# Single-exposure elemental differentiation and texture-sensitive phase-retrieval imaging with a neutron counting micro-channel plate detector

Benedicta D. Arhatari[1,2,]*, David M. Paganin[3,]*, Henry Kirkwood[4], Anton S. Tremsin[5], Timur E. Gureyev[6,3], Alexander M. Korsunsky[7], Winfried Kockelmann[8], Felix Hofmann[7], Eric Huwald[2], Shu-Yan Zhang[9], Joe Kelleher[8], Brian Abbey[2]

[1]*Australian Synchrotron, ANSTO, Clayton, Victoria 3168, Australia*

[2]*Department of Mathematical and Physical Sciences, La Trobe University, Bundoora, Victoria 3086, Australia*

[3]*School of Physics and Astronomy, Monash University, Clayton, Victoria 3800, Australia*

[4]*European XFEL GmbH, 22869 Schenefeld, Germany*

[5]*SSL, University of California, Berkeley, CA 94720, USA*

[6]*School of Physics, The University of Melbourne, Parkville, Victoria 3010, Australia*

[7]*Department of Engineering Science, University of Oxford, Parks Road, Oxford, OX1 3PJ, UK*

[8]*STFC-Rutherford Appleton Laboratory, ISIS Facility, Harwell OX11 0QX, UK*

[9]*Center of Excellence for Advanced Materials, Songshan Lake Industrial Park, Dongguan, Guangdong 523808, China*

* Both authors contributed equally to this paper

**ABSTRACT**

Micro-channel plate (MCP) detectors, when used at pulsed-neutron-source instruments, offer the possibility of high spatial resolution and high contrast imaging with pixel-level spectroscopic information. Here we demonstrate the possibility of multimodal analysis including total neutron cross-section spectra measurements, quantitative material differentiation imaging and texture-sensitive in-line phase imaging, from a single exposure using an MCP detector. This multimodal approach operates in full-field imaging mode, with the neutron transmission spectra acquired at each individual detector pixel. Due to the polychromatic nature of the beam and spectroscopic resolving capability of the detector, no energy scanning is required. Good agreement with the library reference data is demonstrated for neutron cross-section spectra measurements. Two different images corresponding to two selected energy bandwidths are used for elemental differentiation imaging. Moreover, the presence of changes in texture, i.e., preferred grain orientation, in the sample is identified from our phase-retrieval imaging results.

## I. INTRODUCTION

Neutrons offer the possibility of non-destructive imaging and *operando* testing of large samples and components, as well as entire assemblies. The information obtained from neutron experiments is often complementary to x-rays, and many techniques developed for x-rays can be adapted for neutron measurements [1-3] and *vice versa*. Whilst x-rays are scattered and absorbed by the electron cloud of an atom, neutrons interact with the atomic nuclei. In addition, neutrons are more sensitive to the presence of some light elements than x-rays [4]. Contrary to the case of x-rays, neutron attenuation coefficients are not proportional to the atomic number of the elements they interact with [5]. Similar to x-rays though, the degree of neutron beam attenuation is dependent on the incident beam energy. Small-angle scattering, using both neutrons and x-rays, is similarly described by a form factor. Crystal or particle form factors are also important at the energy of slow neutrons, which do not resolve nuclear structure due to the neutron wavelength being about 5 orders of magnitude larger than the nuclear radius [6]. The apparent absorption for neutrons increases with increasing distance between the sample and the detector due to scattering into the field of view dropping off (which is also the case for x-rays). In terms of spatial resolution, x-rays can resolve features down to the nanoscale. However, with a penetration depth on the order of centimeters in high-Z materials, neutrons in many ways constitute the ideal tool for engineering and industrial applications [4].

The probability of neutron-matter interactions is described by the neutron absorption and scattering cross-sections determined by the isotopic composition of the material, the material properties, and the neutron wavelength [7]. The neutron cross-section is an effective area indicating the probability of scattering and absorption for an incident neutron. This neutron cross-section can be measured with high accuracy spectroscopically, for example, using the time-of-flight (TOF) technique [8]. The term 'total neutron cross-section' includes the contributions of absorption, and coherent, incoherent, elastic, and inelastic scattering [9]. The neutron cross-section values are very useful for certain types of analysis, such as elemental differentiation imaging, which is discussed further in this paper.

The elemental specificity of x-rays and neutrons has a wide range of implications and applications [10]. Generating elemental distribution maps using x-rays can be achieved primarily in two different ways: through analysis of the characteristic peaks present in x-ray fluorescence (XRF) spectra [11] and by analyzing the intensity of sample absorption edges [12] (e.g. K-edge subtraction, KES). With neutrons, full-field elemental differentiation imaging can be achieved by exploiting the resonance properties of the neutron cross-section for each individual element of interest [13]. For example, Fast Neutron Resonance Radiography (NRR) has been used to distinguish between light elements such as oxygen, nitrogen, carbon, and hydrogen by exploiting the cross-section resonance features for these elements. Kockelmann et al. [14] demonstrated neutron crystallographic phase contrast radiography by using pairs of wavelengths below and above the characteristic Bragg edge and by using two different types of detectors, namely a 'triggerable' CCD for imaging and a TOF detector for energy selection.

On the other hand, phase imaging using neutrons on the basis of the refractive index [15] has distinctive challenges and a long history of development [16,17]. The difficulties are mainly caused by the fact that phase imaging places strong requirements on the coherence of the neutron sources. In spite of this problem, neutron phase-contrast radiography [17,18] and neutron phase measurements have been successfully demonstrated by measuring intensities at two different sample-to-detector propagation distances [15,16], with more recent work showing that such measurements may also be performed for single-material samples using a single propagation distance [19,20]. Another common technique for phase contrast using neutrons is grating-based interferometry, where diffraction gratings are used to create spatially periodic patterns that can be used to detect sample-induced phase shifts, attenuation, and diffusive dark field signals that are associated with local small-angle scatter [21,22]. A difference between grating-based interferometry and single-propagation-distance-based phase contrast is that the latter technique has a simpler set-up without the need for post-specimen optical elements in order to form a phase-contrast image. However, this simplicity comes with the associated strong cost of a significantly reduced domain of validity, as well as an increased reliance on a sufficient degree of spatial coherence, in comparison to grating-based neutron imaging.

A micro-channel plate (MCP) stack combined with the Timepix detector [23] for readout was used in this study to collect a spatially resolved spectroscopic image of a composite cylindrical sample made from copper and aluminum parts. This MCP detector [24] has proven successful in measuring residual strain with high accuracy [25], by combining transmission radiography with the neutron counting ability of the detector in neutron TOF experiments [26]. It was also used in neutron strain tomography experiments [27-32], to image neutron refraction contrast and edge effects [18], and for neutron micro-tomography [33].

Our multimodal approach utilizes the natural bandwidth of the neutron pulses, without any requirement to scan the incident energy. In this study, we exploit the energy-resolving properties of neutron counting detectors used at pulsed neutron sources. With only a single exposure, we can deliver a three-fold multimodal imaging approach, for (i) total neutron cross-section spectra measurements, (ii) quantitative material differentiation imaging, and (iii) phase-retrieval-enabled texture-sensitive imaging. We now briefly elaborate on each of these three complementary simultaneously acquired imaging channels, that together constitute our multimodal imaging method. Firstly, we can measure the neutron cross section as a function of neutron wavelength and compare it with the theoretically calculated neutron cross section obtained using the *nxs* code [9]. The neutron cross sections, that can be derived from the absorption coefficient data as a function of the neutron wavelength, were measured from the transmitted intensity images and verified using the neutron cross section database of the corresponding element in the sample. Secondly, we can also achieve quantitative elemental differentiation imaging. By solving a standard set of matrix equations based on the Beer-Lambert law, we were able to recover quantitative elemental information. Thirdly, a texture-sensitive phase-retrieval algorithm can be applied. In the absence of the influence of texture, such an analysis would give the retrieved projected thickness for each individual elemental component of the sample. The presence of highly textured regions in a material alters this retrieval, perturbing the recovered signal into an effective projected thickness, or, stated more precisely, an effective projected linear attenuation coefficient that is altered by the presence of the texture. In highly textured materials, grains are oriented in a preferred direction that may be described in the usual manner via an appropriate orientation distribution function [34,35]. This changes neutron coherent scattering in those areas, affects the transmitted intensity, and enables reconstruction of the presence of these changes through neutron phase-retrieval-enabled imaging. A key ingredient, here, is the use of single-plane propagation-based phase retrieval, which can recover a projected linear attenuation coefficient with enhanced signal-to-noise ratio in comparison to analysis based on Beer's Law [19,36]. As shall be shown theoretically and then demonstrated experimentally, this phase-retrieval-enabled approach enables one to recover a texture-sensitive effective projected signal, in the context of our trimodal neutron imaging method.

We close this introduction with a brief overview of the remainder of the paper. Our materials and methods are outlined in Sec. II, which separately considers the sample employed for our neutron-imaging studies (Sec. II A), the experimental setup that was used (Sec. II B), the means utilized for quantitative elemental differentiation (Sec. II C), and our method for texture-sensitive neutron imaging that is enabled by propagation-based phase contrast (Sec. II D). Our results and discussion appear in Sec. III, which is broken into parts that separately consider our neutron cross-section spectra measurements (Sec. III A), elemental differentiation based on attenuation (Sec. III B), and neutron texture-sensitive phase imaging (Sec. III C). Concluding remarks are given in Sec. IV, with an Appendix covering additional mathematical detail regarding texture-sensitive phase-retrieval-enabled propagation-based neutron phase contrast imaging.

## II. MATERIALS AND METHODS

### A. Sample

For this study, we constructed a test sample composed of two cylindrical elemental solids (aluminum and copper). Two samples were placed on top of one another during measurements in order to make the most use of the available neutron flux by filling the incident beam. A photograph of the sample is shown in Fig. 1(a). The outer diameter of the copper cylinder was 5 mm, whilst that of the aluminum cylinder was 14 mm. The aluminum cylinder (made of 6061 *Al* alloy) was an axisymmetric shrink-fitted 'ring-and-plug' assembly and consisted of two separate aluminum parts: a hollow outer cylinder referred to as the 'ring', and a solid inner cylinder, the 'plug'. The ring had inner and outer circle diameters of 7 mm and 14 mm, respectively. The plug was elliptical in shape, with diameters along the major and minor axes of 7.03 mm and 7.00 mm respectively. Prior to assembly of the ring-and-plug specimen, the outer hollow cylinder was heated above 300°C, causing the inner diameter to increase to 7.1 mm and allowing the plug to be inserted freely. The ring was then allowed to cool to room temperature, creating an asymmetric internal stress state.

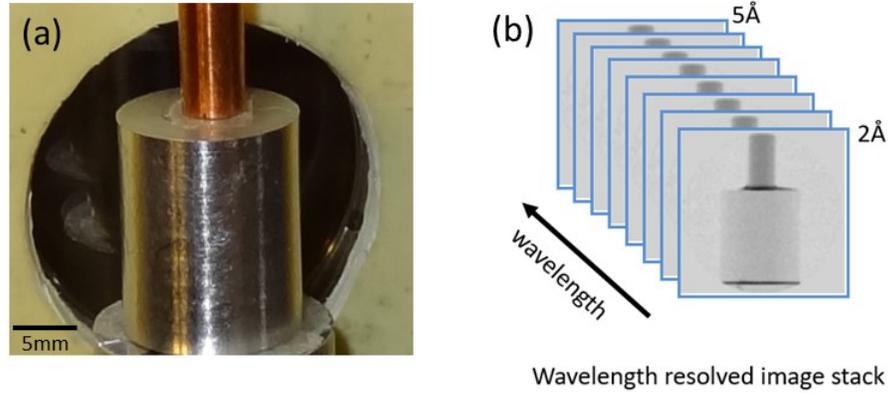

Figure 1. (a) Photograph of the sample. (b) One projection of wavelength-resolved image stack produced by MCP detector.

**B. Neutron radiography with the MCP detector**

The experiment for this study was performed at the ENGIN-X [37] beamline at the ISIS neutron facility, Rutherford-Appleton Laboratory, Harwell Campus, UK. The MCP detector, used in this study, was a combination of spatial imaging and energy discriminating neutron counting detector. Boron was added to the MCP glass structure to convert neutrons into charged particles through the $^{10}B(n,\alpha)^7Li$ reaction [38]. When an incident neutron is captured by a $^{10}B$ atom, an alpha particle and $^7Li$ reaction products are emitted. These energetic charged particles release secondary electrons, initiating an electron avalanche and strong output pulse. The detector consisted of a 512 × 512 matrix of 55 μm square channel geometry or pixels. The energy of each registered neutron was reconstructed by the time-of-flight technique [26]. Neutron transmission spectra were measured in each pixel of this square channel geometry, simultaneously. The number of spectral images can be set between 1 and approximately 3000. In our case, a set of 2031 spectroscopic radiography images were recorded during a single 30-minute exposure for one projection. Thus, one projection image consists of a wavelength-resolved image stack, as shown in Fig. 1(b). The spectral information in each of the radiography images ranged from 2 Å to 5 Å for the neutron wavelength. The MCP-based neutron detector used in this work did not record the neutron wavelength continuously over the entire wavelength range. Rather, it recorded 5 different bands with a 0.25 Å bandwidth for each band, and a wavelength gap of 0.2Å to 0.5Å between each band. The sample was mounted 50 mm in front of the detector and 50 m downstream of the pulsed neutron source. There is no beam collimation for this experiment as we were aiming to illuminate the entire detector. The aperture size over the detector is 28 mm, and the geometric magnification is approximately unity. This flight path allows the neutron pulse to expand sufficiently in time to attain a wavelength resolution of 60 fm when using the MCP detector. In order to study the angular dependence of the sample transmission that arises due to the texture of the sample, 46 equally spaced projections were recorded by rotating the sample through the angular range from 0° to 180°, again with a 30-minute exposure time for each projection.

### C. Quantitative elemental differentiation based on attenuation

Here, we briefly review the standard matrix-inversion method for quantitative elemental differentiation based on attenuation, which is commonly employed for both neutron [39] and x-ray [40-42] studies. Consider two different neutron wavelengths of $\lambda_1$ and $\lambda_2$, with open beam intensities $I_{1,in}$ and $I_{2,in}$, and measured transmitted intensities $I_1$ and $I_2$, respectively. Using the Beer-Lambert law [43], the transmitted intensities can be written in the usual way in terms of the neutron attenuation coefficients $\mu_{1,Cu}, \mu_{1,Al}, \mu_{2,Cu}, \mu_{2,Al}$ (for each specific element and wavelength) and sample thicknesses in terms of each element $t_{Cu}(\mathbf{r}), t_{Al}(\mathbf{r})$, where $\mathbf{r} = (x,y)$ denotes transverse position in the image:

$$I_1(r) = exp[-\mu_{1,Cu}t_{Cu}(r) - \mu_{1,Al}t_{Al}(r)]I_{1,in}(r),$$

$$I_2(r) = exp[-\mu_{2,Cu}t_{Cu}(r) - \mu_{2,Al}t_{Al}(r)]I_{2,in}(r). \quad (1)$$

Here subscripts *Cu* and *Al* denote copper and aluminum respectively and the superscripts *1* and *2* correspond to the wavelengths $\lambda_1$ and $\lambda_2$.

Equation (1) can be written in the standard matrix form at each point *r* as

$$\begin{pmatrix} ln(I_1/I_{1,in}) \\ ln(I_2/I_{2,in}) \end{pmatrix} = \begin{pmatrix} -\mu_{1,Cu} & -\mu_{1,Al} \\ -\mu_{2,Cu} & -\mu_{2,Al} \end{pmatrix} \begin{pmatrix} t_{Cu} \\ t_{Al} \end{pmatrix}, \quad (2)$$

which can be solved as follows:

$$\begin{pmatrix} t_{Cu} \\ t_{Al} \end{pmatrix} = \frac{1}{\mu_{1,Cu}\mu_{2,Al} - \mu_{1,Al}\mu_{2,Cu}} \begin{pmatrix} -\mu_{2,Al} & \mu_{1,Al} \\ \mu_{2,Cu} & -\mu_{1,Cu} \end{pmatrix} \begin{pmatrix} ln(I_1/I_{1,in}) \\ ln(I_2/I_{2,in}) \end{pmatrix}. \quad (3)$$

Equation (3) shows that two transmitted beam intensities measured at two different wavelengths are required to solve for the projected thicknesses of the two elemental distributions, provided that the determinant of the matrix in Eq. (2) (denominator in Eq. (3)) is not close to zero. This matrix-method can be extended, for example if a sample consists of 3 different elements, then we need 3 transmitted beam intensities at 3 different wavelengths.

### D. Neutron texture-sensitive phase imaging

When the projected thicknesses, $t_{Cu}(r), t_{Al}(r)$, of the component materials are recovered as described above, the phase distribution at any wavelength can be calculated using the projection approximation as [36]

$$\varphi(r;\lambda) = -\frac{2\pi}{\lambda}[\delta_{Al}(\lambda)t_{Al}(r) + \delta_{Cu}(\lambda)t_{Cu}(r)]. \quad (4)$$

Here $\delta_{Cu}, \delta_{Al}$ denote the decrement of the real part of the refractive index, which can be calculated for neutrons using the Fermi thin-slab formula [44,45] as

$$\delta(\lambda) = \frac{\lambda^2 \rho_A}{2\pi} b(\lambda). \tag{5}$$

Here, $\rho_A$ is the atomic number density and $b$ is the coherent bound scattering length. Therefore, the phase distribution of Eq. (4) can be written as

$$\varphi(\boldsymbol{r}; \lambda) = -\lambda[\rho_{Al} b_{Al}(\lambda) t_{Al}(\boldsymbol{r}) + \rho_{Cu} b_{Cu}(\lambda) t_{Cu}(\boldsymbol{r})]. \tag{6}$$

Phase measurements can also be made indirectly using a suitable phase retrieval algorithm. There are several phase retrieval approaches that are currently used for propagation-based contrast in the measured intensity. Phase retrieval based on the Transport of Intensity Equation (TIE) [46] is primarily used for quantitative phase computation in the near-Fresnel region. All objects consisting of a single material will here be categorized by the term 'homogeneous' [36]. For such a homogeneous sample, the TIE based phase retrieval (TIE-Hom) [36] algorithm (also known as 'Paganin's method') can be applied for a single measured propagation-based phase-contrast intensity image, $I_z(\boldsymbol{r})$, as follows:

$$t(\boldsymbol{r}) = -\frac{1}{\mu_\lambda} ln\left[\mathbb{F}^{-1}\left(\frac{\mu_\lambda}{\mu_\lambda + z\delta_\lambda |\boldsymbol{u}|^2} \mathbb{F}\left[\frac{I_z(\boldsymbol{r})}{I_{in}}\right]\right)\right]. \tag{7}$$

Here $\mathbb{F}$ is the Fourier transform operator with respect to the transverse coordinates $\boldsymbol{r}$, $\mathbb{F}^{-1}$ is the corresponding inverse Fourier transform, $z$ is the sample-to-detector propagation distance, $\boldsymbol{u}$ is the Fourier variable conjugate to $\boldsymbol{r}$, and $\lambda$ subscripts indicate functional dependence on wavelength. Note that we may employ any Fourier transform convention whereby the Fourier derivative theorem maps the two-dimensional transverse Laplacian operator to multiplication by $-|\boldsymbol{u}|^2$. This phase retrieval method recovers the projected sample thickness with increased signal-to-noise ratio (SNR) relative to the SNR of the measured intensity data [47-49]. The retrieved thickness can then be presented in the form of a phase shift using the one-material version of Eq. (6) [cf. Eq. (A2) in the Appendix]. This particular TIE-Hom method of phase retrieval is used in the present study.

Note, also, that while the algorithm was originally developed in the context of hard x-ray propagation-based phase contrast and phase retrieval, our original x-ray publication [36] indicated that the method could be applied to a variety of radiation and matter wave fields, including neutrons. Indeed, the method has subsequently been applied to both electrons [50-52], and, very recently, to neutrons [19,20]. When recast in terms of key neutron-imaging parameters, the preceding equation takes the equivalent form [19]

$$\rho_A^{(p)}(\boldsymbol{r}) = -\frac{1}{\sigma} ln\left[\mathbb{F}^{-1}\left(\frac{2\pi\sigma}{2\pi\sigma + b\lambda^2 z |\boldsymbol{u}|^2}\mathbb{F}\left[\frac{I_z(\boldsymbol{r})}{I_{in}}\right]\right)\right]. \qquad (8)$$

Here, $\rho_A^{(p)}(\boldsymbol{r})$ is the projected atomic number density (with this projection being taken in the direction of the optical axis $z$), $\boldsymbol{r}$ again denotes spatial coordinates in planes perpendicular to the optical axis, and $\sigma$ is the total neutron cross section. To convert between the language of 'projected thickness' (as employed in a number of x-ray papers that have utilized Eq. (7) in an x-ray context [53]) and the language of neutron optics, we note that, for a sample that is locally composed of a single material,

$$\mu_\lambda t(\boldsymbol{r}) = \sigma \rho_A^{(p)}(\boldsymbol{r}). \qquad (9)$$

While we employ the language of 'projected thickness' in the present paper, the above expression shows this to be proportional to the projected atomic number density.

There are only two published papers [19,20] which have applied Paganin's method to neutrons. However, both of these works ignore the influence of texture upon the recovered projected thickness, in the sense of a position-dependent orientation distribution function [34,35] due to crystallite directions that are not locally completely randomly oriented. From one perspective, we could merely state that Paganin's method breaks down in the presence of sample texture. However, given the extreme SNR-boosting property of the method—which has enabled, for example, image acquisition-time reductions of over three orders of magnitude in the x-ray domain [54], thereby enabling over one thousand x-ray tomograms per second to be recovered [55,56]—there is strong motivation to adapt the neutron variant of the method to account for the presence of texture. As additional context, Paganin's method has recently been adapted to a different form of unresolved spatially-random microstructure, associated with unresolved random refractive-index variations in an x-ray setting [57,58], which opens the logical possibility that the neutron version of the method may be adapted to take texture into account.

Suppose, then, that a single propagation-based phase-contrast neutron image is taken of a homogeneous textured sample, with this image $I_z(\mathbf{r})$ being processed using the Paganin filter according to Eq. (7). Introducing the linear operator

$$\widehat{\Phi} = \mathbb{F}^{-1} \frac{\mu_\lambda}{\mu_\lambda + z\delta_\lambda |\mathbf{u}|^2} \mathbb{F} \frac{1}{I_{in}} , \qquad (10)$$

wherein all operators act in the order from right to left, this process may be denoted by the nonlinear operation [36]

$$t_{eff}(\mathbf{r}) = -\frac{1}{\mu_\lambda} \ln \widehat{\Phi} I_z(\mathbf{r}) . \qquad (11)$$

The theoretical value of $\mu_\lambda$ is calculated for a random powder (non-textured). Here, the influence of sample texture implies that the left side of the above expression, rather than being the actual projected thickness $t(\mathbf{r})$, is instead a texture-sensitive effective projected thickness $t_{eff}(\mathbf{r})$. As shown in the Appendix, the effective projected thickness is related to the actual projected thickness via

$$t_{eff}(\mathbf{r}) = t(\mathbf{r}) + C(\mathbf{r}). \qquad (12)$$

While the above expression may be taken as amounting to a definition of the texture-sensitive correction term $C(\mathbf{r})$, the Appendix shows how this may be written as

$$C(\mathbf{r}) = -\frac{1}{\mu_\lambda}\ln\left[e^{-\mu_\lambda t(\mathbf{r})} + \widehat{\Phi}\mathcal{A}(\mathbf{r})\right] - t(\mathbf{r}). \qquad (13)$$

Here, the explicitly texture-sensitive term $\mathcal{A}(\mathbf{r})$ is given by (omitting functional dependence on $\mathbf{r}$ for brevity)

$$\mathcal{A} = \mathbb{E}[\tilde{I}] - \frac{z}{k}\nabla_\perp \cdot (\mathbb{E}[I_{in}\nabla_\perp\tilde{\varphi} + \tilde{I}\nabla_\perp\varphi + \tilde{I}\nabla_\perp\tilde{\varphi}]). \qquad (14)$$

In the preceding expression, $\nabla_\perp$ denotes the transverse gradient, $k = \frac{2\pi}{\lambda}$ denotes the wavenumber corresponding to the wavelength $\lambda$, $\mathbb{E}[\ ]$ denotes an expectation value with respect to an ensemble of statistical realizations of the spatially random sample texture, and for any member of this statistical ensemble we write

$$\begin{aligned} I_{z=0} &\rightarrow I_{z=0} + \tilde{I} \equiv I'_{z=0} \\ \varphi &\rightarrow \varphi + \tilde{\varphi} \equiv \varphi' \end{aligned} \qquad (15)$$

in order to quantify the influence of texture (together with other rapidly-spatially-varying stochastic fluctuations that are below the resolution limit of the imaging system, should they be present). Note, in this context, that (i) $\varphi \equiv \varphi(\mathbf{r})$ denotes the phase function that would exist in the absence of textural influence, (ii) the stochastic quantities $\tilde{I}$ and $\tilde{\varphi}$ both depend on transverse position $\mathbf{r}$, (iii) the expectation value of each of these stochastic quantities does not necessarily vanish, but even when this is the case, the covariance term $\mathbb{E}[\tilde{I}\nabla_\perp\tilde{\varphi}]$ in Eq. (14) will typically not vanish, (iv) both stochastic quantities ($\tilde{I}$ and $\tilde{\varphi}$)—together with the auxiliary function $\mathcal{A}$ that is derived from them—will depend on the sample texture and sample orientation, (v) the influence of $\tilde{I}$ is likely to be considerably stronger than that of $\tilde{\varphi}$ in the present neutron-texture context, and (vi) one possible physical mechanism for nonzero $\tilde{\varphi}$, if present, might be due to orientational disorder associated with texture implying an associated disorder in the exit-surface neutron wavefront, with the latter quantity containing rapidly spatially varying zero-average spatial fluctuations in the projected thickness $t$, as a form of "phase roughness" associated with the projection of a large number of crystallites.

In the absence of sample texture, $\tilde{I}$ and $\tilde{\varphi}$ in Eq. (15) both vanish, so that $\mathcal{A}$ (as defined in Eq. (14)) also vanishes in the zero-texture limit. The texture-dependent correction $C$ in Eq. (13) then vanishes, with Eq. (12) thereby indicating that the effective projected thickness, $t_{eff}$, reduces to the texture-free projected thickness $t$.

As this summary of the mathematical derivation in the Appendix shows, when a textured sample is employed to form a propagation-based phase contrast neutron image $I_z(\boldsymbol{r})$ that is subsequently processed using the Paganin filter in the form given by Eq. (7), the recovered effective projected thickness (see Eq. (12)) is equal to the actual projected thickness $t(\boldsymbol{r})$ plus a texture-sensitive correction $C(\boldsymbol{r})$ (see Eqs. (12)-(15)). This correction will be manifest as an orientation-dependent signal, in the recovered effective projected thickness. Moreover, since texture serves to increase the effective linear attenuation coefficient, the correction $C(\boldsymbol{r})$ will typically be positive. Finally, we point out that the effective projected thickness may differ considerably from the actual projected thickness, e.g. due to the influence of texture on the attenuation of a neutron beam. We emphasize that, when such a considerable difference exists, it is indicative of a strong deviation of the scattering cross section from the texture-free limit, rather than being suggestive of unphysically strong compressive strain.

## III. RESULTS AND DISCUSSION

### A. Neutron cross-section spectra measurement

In this study, we compared the experimental and theoretical neutron cross-section spectra for the Al and Cu samples respectively. The measured transmitted neutron intensity was normalized with respect to the open beam intensity to determine the experimental attenuation coefficient at the known object thickness, $t$:

$$\mu(\lambda) = -\frac{ln(I(\lambda)/I_{in})}{t}. \tag{16}$$

The measured transmitted intensity in a region located in the center of the sample was used, with $t$ set to the sample diameter (5 mm for copper and 14 mm for aluminum respectively). The theoretical transmission spectra were calculated from the theoretical total cross section, $\sigma(\lambda)$, from the *nxs* program library [9]. The relationship between the attenuation coefficients and the cross-section is [9]

$$\mu(\lambda) = N\rho_A \sigma(\lambda), \tag{17}$$

where $N$ is the number of atoms per unit cell and $\rho_A$ is the atomic number density. The comparison of experimental and theoretical predicted values of the attenuation coefficient as a function of wavelength is shown in Fig. 2(a) and 2(b) for copper and aluminum respectively. Figure 2 also shows the theoretical characteristic case of Bragg edges for both *Cu* and *Al* powder (non-textured). The MCP detector used in this work recorded 5 distinguishable bands, as can be seen in the experimental data for copper and aluminum (Fig. 2(a) and 2(b)). In order to reduce the fluctuation of the experimental data, a smoothing filter was applied with a width of 15 channels. The experimental attenuation coefficient result for *Cu* can be seen in Fig. 2(a). The discrepancy between the theoretical and experimental values for *Cu* is represented by a chi-squared value of 0.0096 [59]. The theoretical values for *Al* in Fig. 2(b) correspond to the texture-free case (*Al* powder). These values coincide with those generated by Boin [9], for untextured *Al*. However, the result in Boin's article was added by simulation using the March-Dollase model [60] to include *Al* with preferred crystallographic orientation (texture). This March-Dollase model simulation result for textured *Al* is in good agreement with our experimental results for *Al* in Fig. 2(b). This shows that our sample is indeed textured *Al*. A signature of textured material is usually indicated by the manner in which the absorption coefficient changes before the edge, and on the height of the Bragg edges such as a peak in the (111) aluminum edge. In our experiment with an uncollimated neutron beam, we consider that modeling the linear attenuation as that of pure Al powder is an acceptable approximation, even though the sample is in fact an Al alloy. The alloying elements can be important for the scattering of low-energy neutrons [61], where alloy precipitates contribute to the small-angle scattering background.

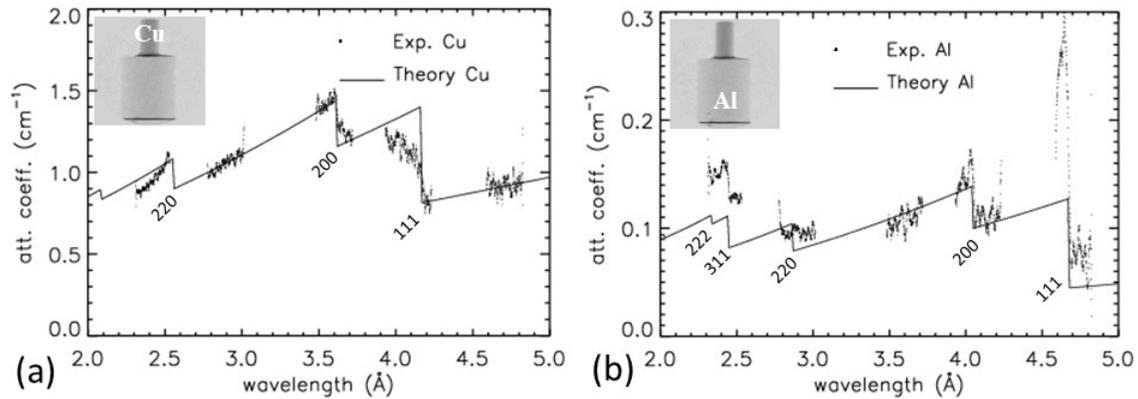

Figure 2. A comparison of the experimental values of attenuation coefficient spectra with the theoretical (untextured material) values from the *nxs* library for (a) copper and (b) aluminum. The experimental result is shown for 5 different energy bands.

## B. Quantitative elemental differentiation based on attenuation

Two different narrow-wavelength neutron bandwidths were selected in order to perform elemental differentiation (in this case these elements are not mixed to form a new compound) and extract phase information. The first bandwidth centered at 2.4 Å spanned a total range of 0.09 Å comprising 147 radiography images (from a total of 2031 images in one data set ranging from 2 Å to 5 Å), whilst the second bandwidth centered at 2.5 Å spanned a range of 0.08 Å and comprised 128 radiography images. The plot of the theoretical attenuation coefficient of *Al* and *Cu* powder (both free from texture and strain) in this selected wavelength bandwidth is shown in Fig. 3. The first bandwidth is represented in blue, while the second bandwidth is in yellow. Both wavelength bands show regions of linear behavior of attenuation values that do not include the Bragg edge position.

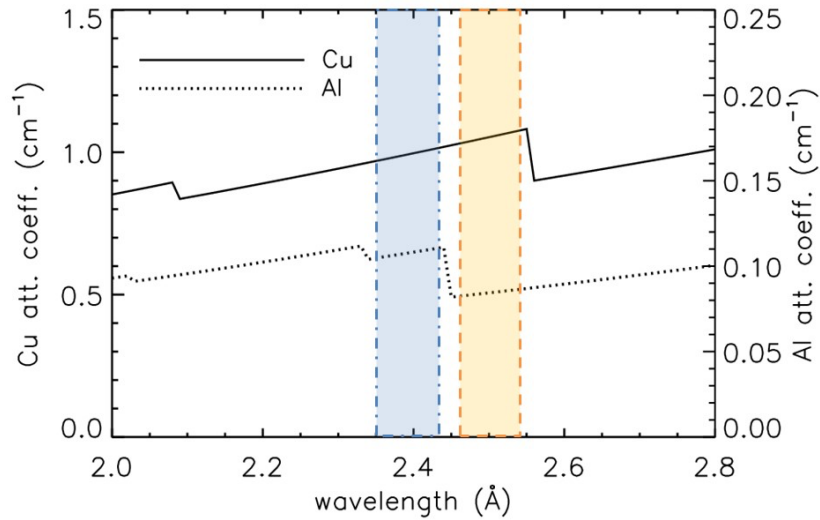

Figure 3. Theoretical attenuation coefficient spectra for *Cu* and *Al* within the two selected wavelength bands. The first bandwidth (centered at 2.4 Å) is depicted in blue, while the second bandwidth (centered at 2.5 Å) is depicted in yellow.

Within these narrow wavelength bands, there is a linear relationship between the wavelength and the attenuation coefficient. To enable averaging of the intensities in these narrow wavelength bands for the purpose of improving signal-to-noise ratio, we make the assumption of a "thin" sample, so that $\mu_\lambda t \ll 1$. In this way, we can average the intensity radiographic images of the sample within each band and representing each band using a single average wavelength. The average intensity radiographs, $\overline{I}_1$ and $\overline{I}_2$, for the two bandwidths are shown in Fig. 4(a) and 4(b) after normalization by the open beam intensity. Note that the thin dark regions at the top and bottom of the aluminum cylinder correspond to the adhesive used to attach the aluminum cylinder to the base and the copper cylinder to the aluminum (since, as mentioned previously, neutrons exhibit high attenuation for light elements [4] present in glue). Both illumination-corrected images are then used to evaluate Eq. (3) using the theoretical values for the attenuation coefficient of *Cu* and *Al* at wavelengths $\overline{\lambda}_1$ and $\overline{\lambda}_2$, respectively, that are in the middle of the corresponding wavelength bands. The retrieved thicknesses $t_{Cu}$ and $t_{Al}$ are shown in Fig. 4(c) and 4(e) with the corresponding quantitative values plotted in Fig. 4(d) and 4(f). We can see that *Cu* and *Al* are separated in each figure. The plot in Fig. 4(d) has been produced by averaging 100 pixel rows to increase the signal-to-noise ratio, while the plot in Fig. 4(f) averages 200 rows. Similar averaging to increase the signal-to-noise ratio has been applied in the results of the remainder of this paper. The retrieved quantitative thickness profiles accurately reflect the sample sizes, with the outer diameter of the *Cu* cylinder equal to 4.9 ± 0.5 mm, and that of the *Al* cylinder equal to 14 ± 2 mm.

    The matrix solution in Eq. (3) reduces the signal-to-noise ratio (SNR) in the retrieved images of the thickness distribution due to the arithmetic processing in the matrix solution. Stated more explicitly, the SNR reduction arises from the image subtraction implicit in Eq. (3): noise will add under such (weighted) subtractions, while the corresponding signals will

be subtracted, with the net effect being a reduction in SNR [62]. The same conclusion can be reached, and an estimate for the SNR reduction obtained, by considering the condition number of the matrix of coefficients in Eq. (3): the SNR reduction corresponds to a multiplicative factor that is a monotonic function of the ratio of the maximum and minimum eigenvalues of the matrix, since this is a measure of the relative amplification of the most-amplified and least-amplified eigenvectors in the process of solving Eq. (2) to give Eq. (3).

## C. Neutron texture-sensitive phase imaging

With the retrieved thickness profiles presented above, the phase distribution can be calculated at any wavelength using Eq. (6). For the chosen wavelength of 1.798 Å (corresponding to the tabulated coherent bound scattering length [44,45], *b*, at 1.798 Å), the result is presented in Fig. 5. It is interesting to note that *Cu* and *Al* do not show a significant contrast difference in the phase image. Note, also, that since the neutrons have an insufficient degree of coherence to permit interferometric phase determination without additional filtration, the 'phase distribution' referred to above should be understood in terms of the generalized definition of the phase of a partially coherent field in the paper by Paganin and Nugent [63]. Here, rather than viewing phase in terms of the formation (at least in principle) of interference fringes, phase is instead envisaged in terms of the scalar potential associated with a time-averaged or ensemble-averaged current density. The latter construction remains well defined when the field is partially coherent. This can be also interpreted in terms of the "generalized phase" or "generalized eikonal" of a polychromatic wave. Here, the "generalized" quantities are equal to a weighted sum (over the spectrum)

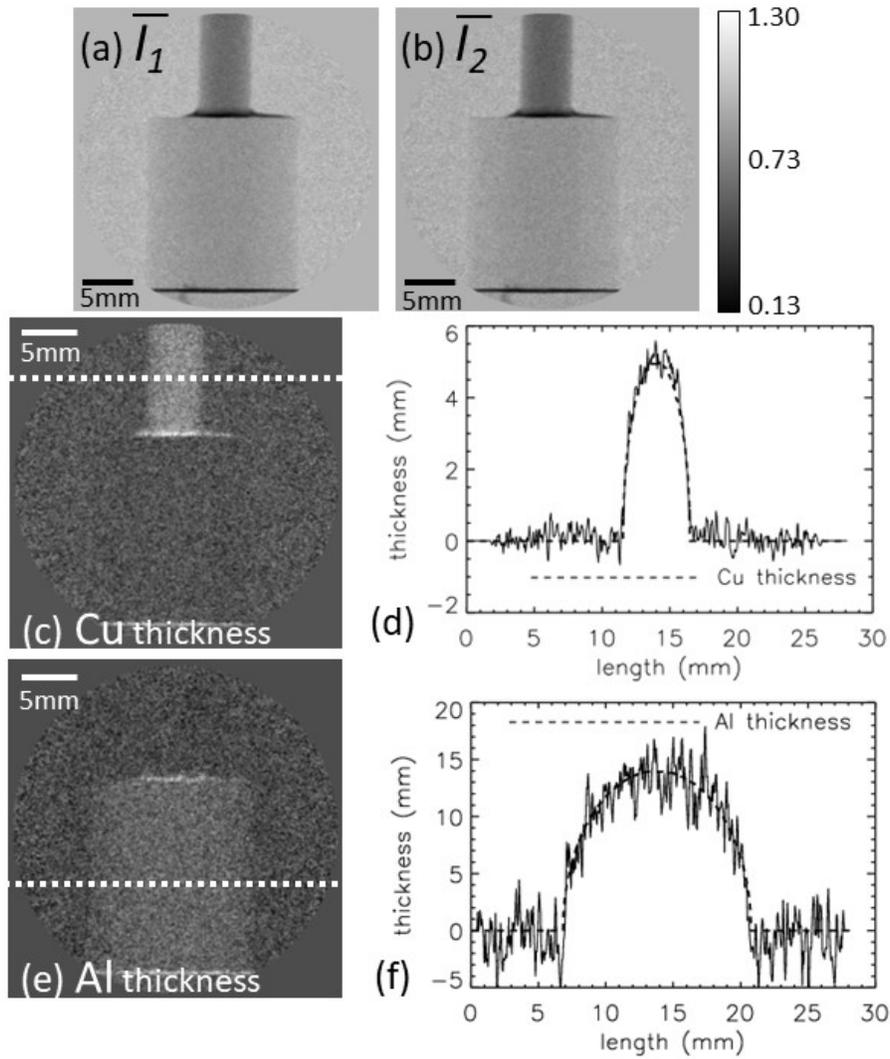

Figure 4. (a) The average intensity distribution at the first bandwidth, $\bar{I}_1$. (b) The average intensity distribution at the second bandwidth, $\bar{I}_2$. (c) The retrieved $Cu$ thickness, $t_{Cu}$, with the corresponding dotted line indicating the lineout taken across the $Cu$ distribution shown in (d). (e) The retrieved $Al$ thickness, $t_{Al}$, with the corresponding dotted line indicating the lineout taken across the $Al$ distribution shown in (f). Dashed lines in (d) and (f) correspond to the theoretical profiles of the projected thickness of $Cu$ and $Al$ respectively.

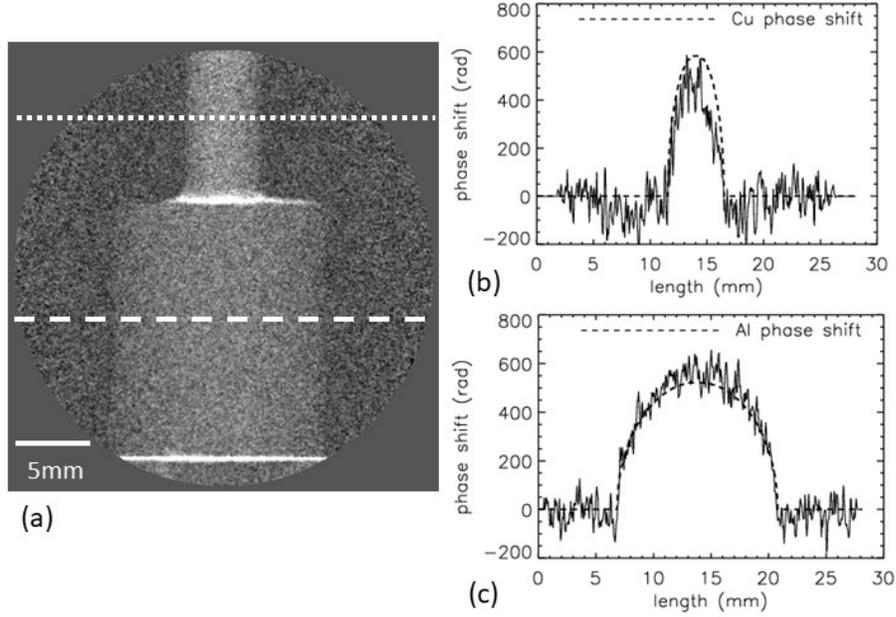

Figure 5. (a) The phase distribution within the sample calculated using Eq. (6) with (b) showing a lineout taken across the *Cu* distribution (corresponding to the dotted line in panel (a)), while (c) shows a lineout taken across the *Al* distribution (corresponding to the dashed line in panel (a)). Dashed lines in (b) and (c) correspond to the theoretical profiles of the projected phase shift of *Cu* and *Al* respectively.

of conventional (coherent) phases or eikonals of monochromatic (fully coherent) components [42].

The TIE-based phase-retrieval algorithm in Eq. (7) can be used to obtain the thickness distribution for each individual element of the sample independently, with improved SNR [47-49]. To fulfil the requirement of the homogeneous sample, firstly, parts of the image pertaining to *Al* and *Cu* were separated. Then, the corresponding values of $\delta$ and $\mu$ for *Cu* and *Al* were used in the TIE-Hom phase retrieval algorithm for homogeneous samples given in Eq. (7) [36]. The retrieved images were then merged back together. We used the averaged intensity data from the wavelength band of 2.5 Å, i.e. $\bar{I}_2$, in Fig. 4(b) as the input for Eq. (7). The resulting retrieved effective thickness, $t_{eff}$, is presented in Fig. 6. Similar results can be obtained by using phase retrieval techniques for multi-material objects [47]. The result presented in Fig. 6 shows improved SNR, as expected [19,47-49]. More importantly, after phase retrieval it was possible to observe an additional apparent thickness contribution, which is directly related to the distinction between thickness and effective thickness that was explained earlier in the paper, in the middle of the sample around the area of the plug. We interpret this additional contrast $C$ as arising from the presence of texture in the sample, which is visible in the retrieved effective thickness (see Eqs. (12)-(15)). Regarding the considerable difference between the actual projected thickness and the effective projected thickness that is evident in Fig. 6(c), it is important to be clear that this is not due to the presence of an unphysically large degree of compressive strain, but rather is due to the considerable influence of texture upon the neutron scattering cross section (cf. the final sentences of Sec. II). The texture present in the sample also results in a higher attenuation coefficient for the sample being measured at 2.5 Å in Fig. 2(b). By applying phase retrieval these texture effects lead to a significant difference in the effective thickness compared to the theoretical thickness profile for the *Al* cylinder.

In order to further investigate the changes in the textural structure of the sample at different sample orientations, Eq. (7) was then applied to each of the 46 projection intensity images ranging from 0º to 180º. We observe that, as expected from the theoretical development in the Appendix, the texture present in the *Al* results in significant changes in the corresponding retrieved effective thickness profile at different rotational positions of the sample, as shown in Fig. 7(a). Figure 7(b) shows the effective thickness profile as a function of sample rotation at 2.5 Å. We observe that there is a small shift between 0º and

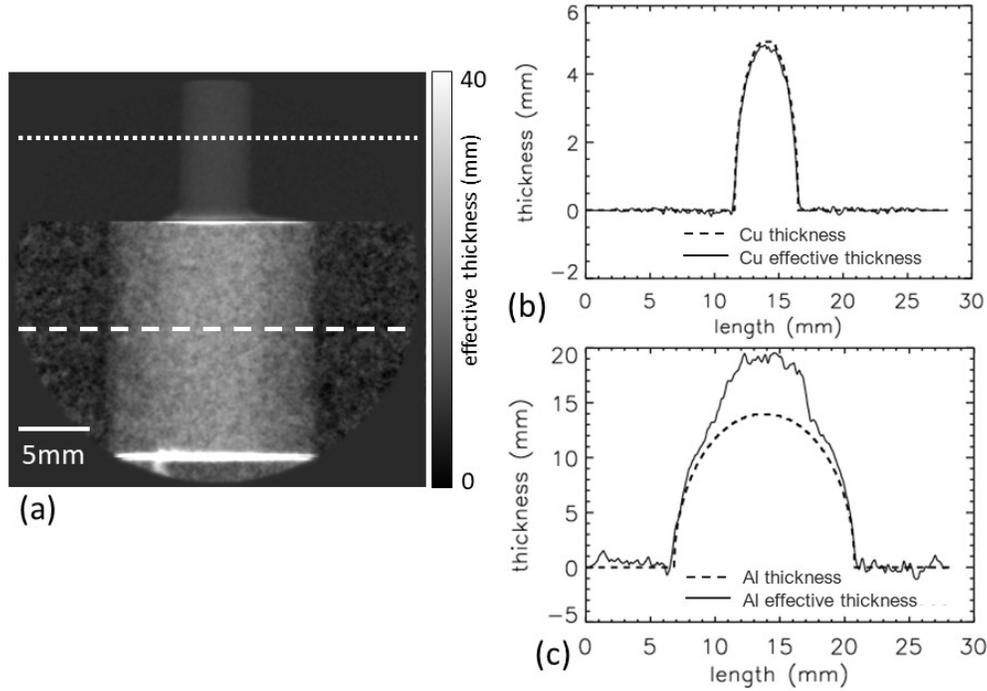

Figure 6. (a) The effective thickness distribution within the sample retrieved using the TIE-Hom algorithm in Eq. (7) for *Al* and *Cu* cylinders independently, with (b) the corresponding lineout taken across the *Cu* distribution (corresponding to the dotted line in panel (a)), and (c) lineout taken across the *Al* distribution (corresponding to the dashed line in panel (a)). Solid lines in (b) and (c) indicate the projected effective thickness of *Cu* and *Al* cylinders respectively. Dashed lines in (b) and (c) indicate the projected thickness of *Cu* and *Al* cylinders, respectively.

180º in position due to the sample precession with rotation. It is important to note that during analysis of these results we determined that the observed effect in the phase retrieved images does not arise from lattice strain which would result in far smaller deviations than can be measured here. This is because strain in the ring-and-plug assembly does not significantly change the neutron transmission between Bragg edges. Additionally, we note that the textural structure in the sample is less observable at a larger wavelength beyond the last Bragg edge [26]. This is consistent with the fact that past the last Bragg edge, texture should have minimal impact on the image contrast in this regime. In this case, we applied the phase retrieval algorithm to the data at 4.75 Å. The resulting data were averaged for every 4 projections, as shown in Fig. 7(c), in order to account for the lower neutron flux at longer wavelengths. Figure 7(c) shows the retrieved effective thickness for several sample rotations. The results confirm that there is no significant variation in retrieved effective thickness as a function of sample rotation at 4.75 Å. This highlights the presence of textural structure effects which are observed at 2.5 Å but not at 4.75 Å.

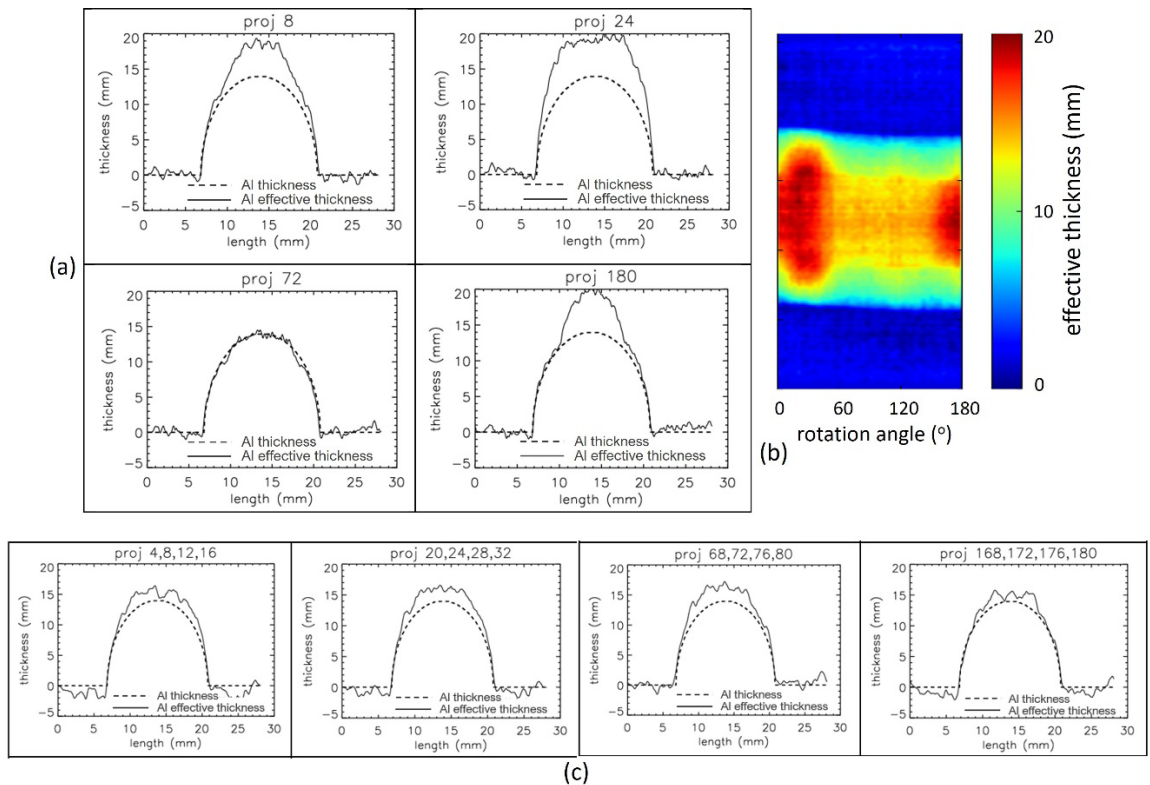

Figure 7. (a) Deviation in the retrieved effective thickness at 2.5 Å determined using the TIE-Hom phase-retrieval algorithm across the *Al* cylinder for 4 rotation angles of the sample from the total of 46 rotation positions. (b) Effective thickness at 2.5 Å for all 46 projections shows the variation as a function of sample rotation. (c) Effective thickness at 4.75 Å does not show any variation as a function of sample rotation.

This textural structure in the grain orientation can also be observed clearly as a function of neutron wavelength because texture is expected to significantly alter the measured neutron transmission at a given wavelength. This texture manifests more strongly in the projected transmission spectrum in the middle of the sample where the ring and plug are superimposed, see Fig. 8(a), in comparison with the area on the ring, see Fig. 8(b). Figure 8 shows the transmission spectra of aluminum at 0° and 90° rotation. The large difference in measured transmission spectra between 0° and 90° rotation is due to texture. This can also be seen when comparing the height of the Bragg edges and the manner in which the transmission changes before the edge. The transmission spectrum measured at the zero-degree projection has typical features that are characteristic of a highly textured sample. Finally, we note that all rolled aluminum materials with highly textured microstructure will show this type of behavior [64]: for example, the disappearance of the 220 Bragg edge and highly non-linear decay of the transmission between 200 and 111 edges at 0° rotation are typical for textured aluminum materials.

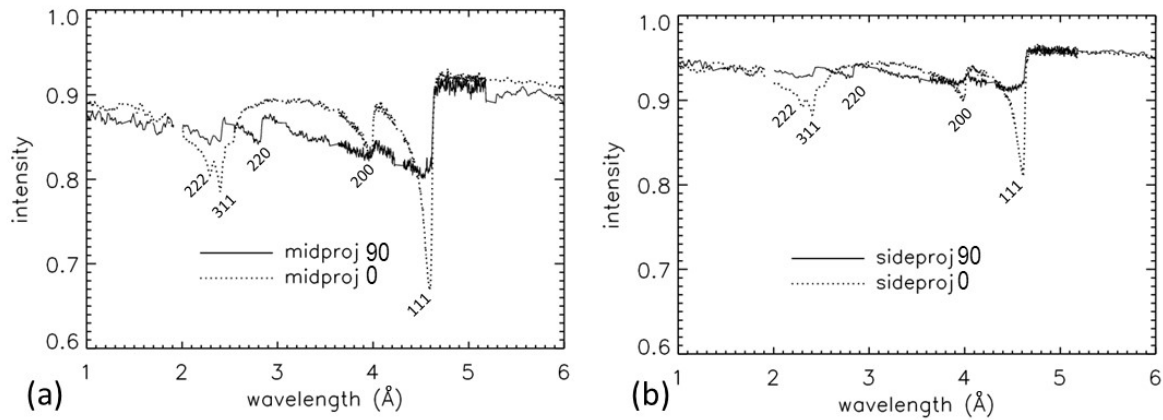

Figure 8. Neutron transmission spectra of *Al* located (a) in the middle of the sample where the ring and plug are superimposed, and (b) on the side of the sample within the ring only. The dotted line is the transmission spectrum at 0°, while the solid line is the transmission spectrum at 90°.

## IV. CONCLUSIONS

We carried out a proof-of-principle multimodal approach demonstrating the benefits of using the energy-resolving properties of neutron counting MCP detectors at pulsed neutron sources. While MCPs have been used for many years for wavelength resolved studies, we stress here the three-fold multimodal information obtainable from a single experimental dataset. Neutron absorption coefficient measurements from experimental data, as a function of neutron wavelength, for textured samples were found to be in good agreement with the simulation result in Boin's article [9]. Spatially resolved spectroscopic information was obtained in one simultaneous measurement using a neutron counting MCP detector and pulsed beam, allowing full-field imaging of the sample at multiple neutron wavelengths without the need to scan through neutron wavelengths. The spectroscopic nature of the data allowed us to apply a standard matrix-based method to solve for the quantitative elemental differentiation of *Cu* and *Al* materials in the sample. This allows efficient non-destructive and quantitative neutron imaging to be performed, using different neutron wavelengths present in the incident pulsed beam. A texture-sensitive phase-retrieval algorithm was also applied, enabling us to retrieve the effective thickness at a neutron wavelength prior to the last Bragg edge. This texture-sensitive phase retrieval method recovers the projected sample effective thickness with significantly increased SNR relative to the SNR of the measured intensity data. The samples used in our study did not exhibit strong texture, as seen in the narrow-band transmission images in our data set. To reconstruct the sample thickness from the measured transmission images, we assumed a texture-free material and used the neutron attenuation coefficient for aluminum powder in the phase-retrieval analysis. The presence of highly textured regions in the *Al* sample can be clearly observed from the texture-sensitive phase-retrieved images. After the last Bragg edge, texture should have minimal impact on the neutron transmitted intensity. This textural structure in the sample changes with different sample orientation, as seen by investigating the retrieved effective thickness of the *Al* sample at different rotation angles between 0º to 180º. The largest difference in the effective thickness was observed at 90º separation. This textural structure in the grain orientation has also been demonstrated in the projected transmission spectrum as a function of neutron wavelength because texture is anticipated to significantly change the measured neutron transmission at a specified wavelength. In the projected transmission spectrum, this texture has a more significant influence in the middle of the sample where the ring and plug are superimposed.

## ACKNOWLEDGEMENTS

The authors acknowledge the support of the Australian Research Council through the Centre of Excellence for Advanced Molecular Imaging.

## APPENDIX: DERIVATION OF TEXTURE-SENSITIVE PHASE-RETRIEVAL IMAGING

In this Appendix, we derive the core results pertaining to texture-sensitive phase-retrieval imaging, that are employed in the main text of the paper. Begin with the following finite-difference form of the transport-of-intensity equation [46]

$$I_z(\boldsymbol{r}) \approx I_{z=0}(\boldsymbol{r}) - \frac{z}{k}\nabla_\perp \cdot [I_{z=0}(\boldsymbol{r})\nabla_\perp \varphi(\boldsymbol{r})], \tag{A1}$$

wherein all symbols are as defined in the main text. This expression quantifies propagation-based phase contrast, in the small-propagation-distance regime corresponding to a Fresnel number that is much greater than unity.

Under the projection approximation, and in the absence of texture-related effects, we may write [36]

$$\varphi = -k\delta_\lambda t, \tag{A2}$$

$$I_{z=0} = I_{in} \exp(-\mu_\lambda t). \tag{A3}$$

Here, explicit functional dependence on $r$ has been suppressed for brevity, and a thin homogeneous sample has been assumed to lie immediately upstream of the plane $z = 0$. Moreover, this sample is assumed to be illuminated by $z$-directed plane waves, of uniform incident intensity $I_{in}$.

In the presence of texture-related effects, we consider both $\varphi$ and $I_{z=0}$ to be perturbed in the manner indicated by Eq. (15) in the main text. The associated perturbations, namely $\tilde{\varphi}$ and $\tilde{I}$, may be interpreted as follows. (i) $\tilde{\varphi} \equiv \tilde{\varphi}(r)$ denotes the perturbations to the phase of the exit-surface neutron field over the plane $z = 0$, due to the influence of sample texture, for any one member in a statistical ensemble of stochastic realizations for that sample texture; (ii) $\tilde{I} = \tilde{I}(r)$ denotes the perturbations to the intensity of the exit-surface neutron field, again due to the influence of sample texture.

To proceed further, substitute Eq. (15) into Eq. (A1), and then take a statistical average over the ensemble of realizations of the sample texture. This gives

$$I_z \approx I_{z=0} - \frac{z}{k}\nabla_\perp \cdot [I_{z=0}\nabla_\perp \varphi] + \mathcal{A}, \tag{A4}$$

where the texture-sensitive auxiliary function $\mathcal{A}$ is as given in Eq. (14) of the main text. Following Paganin et al. [36], Eqs. (A2) and (A3) can be substituted into Eq. (A4), to give

$$I_z \approx I_{in}\left(1 - \frac{z\delta_\lambda}{\mu_\lambda}\nabla_\perp^2\right)e^{-\mu_\lambda t} + \mathcal{A}. \tag{A5}$$

In the zero-texture limit where $\mathcal{A}$ vanishes, this equation reduces to

$$I_z \xrightarrow{\mathcal{A}\to 0} I_{in}\left(1 - \frac{z\delta_\lambda}{\mu_\lambda}\nabla_\perp^2\right)e^{-\mu_\lambda t}. \tag{A6}$$

The linear operator $\widehat{\Phi}$, as given in Eq. (10) from the main text, may be applied to both sides of Eq. (A6) to give

$$e^{-\mu_\lambda t} = \widehat{\Phi} I_z, \qquad \mathcal{A} \to 0. \tag{A7}$$

If we instead apply $\widehat{\Phi}$ to both sides of the texture-sensitive expression in Eq. (A5), we obtain

$$\widehat{\Phi} I_z = e^{-\mu_\lambda t} + \widehat{\Phi}\mathcal{A} \equiv e^{-\mu_\lambda t_{eff}}. \tag{A8}$$

Note that the final part of the preceding expression serves to define the effective projected thickness $t_{eff}$, namely the effective thickness that is produced when the phase-retrieval method of Paganin et al. [36] is applied to the texture-sensitive propagation-based phase-contrast image $I_z$.

To proceed further, take the natural logarithm of the latter part of Eq. (A8), and then make use of Eq. (12) in the main text. We immediately arrive at Eq. (13) of the main text.